# Machine Reading of Hypotheses for Organizational Research Reviews and Pre-trained Models via R Shiny App for Non-Programmers

Victor Zitian Chen
Belk College of Business
University of North Carolina at Charlotte
9201 University City Blvd
Charlotte, NC 28277
Email: zchen23@uncc.edu
Tel: 980 800 1123
**\*Corresponding author**

Wlodek Zadrozny
College of Computing and Informatics
University of North Carolina at Charlotte
9201 University City Blvd
Charlotte, NC 28277
Email: wzadrozn@uncc.edu

Felipe Montano Campos
The Comparative Health Outcomes, Policy, and Economics (CHOICE) Institute
University of Washington at Seattle
1959 NE Pacific Street
Box 357630
Seattle, WA 98195-7630
Email: fmontano@uw.edu

Evan Canfield
Allstate
10800 Sikes Pl
Charlotte, NC 28277
Email: canfielder@gmail.com

# Machine Reading of Hypotheses for Organizational Research Reviews and Pre-trained Models via R Shiny App for Non-Programmers

**Abstract**

The volume of scientific publications in organizational research becomes exceedingly overwhelming for human researchers who seek to timely extract and review knowledge. This paper introduces natural language processing (NLP) models to accelerate the discovery, extraction, and organization of theoretical developments (i.e., hypotheses) from social science publications. We illustrate and evaluate NLP models in the context of a systematic review of stakeholder value constructs and hypotheses. Specifically, we develop NLP models to automatically 1) detect sentences in scholarly documents as hypotheses or not (Hypothesis Detection), 2) deconstruct the hypotheses into nodes (constructs) and links (causal/associative relationships) (Relationship Deconstruction ), and 3) classify the features of links in terms causality (versus association) and direction (positive, negative, versus nonlinear) (Feature Classification). Our models have reported high performance metrics for all three tasks. While our models are built in Python, we have made the pre-trained models fully accessible for non-programmers. We have provided instructions on installing and using our pre-trained models via an R Shiny app graphic user interface (GUI). Finally, we suggest the next paths to extend our methodology for computer-assisted knowledge synthesis.

# INTRODUCTION

Knowledge accessibility is a significant constraint in synthesizing the scientific literature in organizational research (Chen & Hitt, 2021; Larsen, Hekler, Paul, & Gibson, 2020; Li, Larsen, & Abbasi, 2020). A scientific study typically starts with a systematic review of the existing literature, extracting and connecting the published causes-and-effects relationships among constructs of interest. The information extraction work is recognized widely as one of the most challenging and time-consuming activities for research reviews (Felizardo & Carver, 2020). The volume of scientific publications is exceedingly overwhelming for human researchers to synthesize the existing knowledge timely (Antons, Breidbach, Joshi, & Salge, 2021). For instance, a keyword search of "organizational performance" in Web of Science generated about 9,000 papers between 1980-2020, half of which were published in the last five years alone.

Researchers often have to spend limited resources and professional time on tedious manual work of knowledge detection and extraction, yet these efforts may not be sufficiently thorough and timely. It is thus no surprising that recently Antons et al. (2021) call for accessible new methods of computational literature reviews (CLRs) for organizational researchers. They suggest that new methods and tools are needed to engage machine learning algorthms to automatically extract and analyze the content of the text corpus, rather than topics, effect sizes, meta-informationm, or bibliometric analysis (Antons et al., 2021).

While significant advances have been made in recent years in the field of natural language processing (NLP) to train computers to read and comprehend textual data (e.g., OpenAI's `GPT-3`) [for a review, see, e.g., Zhang, Yang, Li, and Wang (2019)], there have been limited developments of NLP models to solve the knowledge inaccessibility problem in reviewing the theoretical content of social science papers. Several efforts were made outside social sciences to extract findings, hypotheses, and descriptive information from scientific publications to assist systematic reviews (Felizardo & Carver, 2020). However, these models are typically built on pre-trained language representations by domain experts and have limited generalizability outside the specific domains where they are developed. So far, almost all the machine reading models for systematic reviews have been developed in biomedicine (Jonnalagadda, Goyal, & Huffman, 2015; Valenzuela-Escárcega et al., 2018). Despite a growing interest in such tools by social and organizational researchers (Chen & Hitt, 2021; Larsen et al., 2020; Li et al., 2020), the development of machine reading models for literature reviews in social sciences, especially in organizational research, has been profoundly limited. The current approaches of computational literature reviews focus primarily on topic modeling and sentiment analysis (for a review, see Antons et al., 2021).

The purpose of this research is thus to introduce to organizational researchers interpretable machine reading approaches to reading and organizing theoretical insights from organizational research papers. We develop NLP models to accelerate the detection, classification, and deconstruction of hypotheses from organizational research publications. This paper, to our knowledge, represents the first efforts to develop machine-aided techniques for theoretical knowledge extraction from scientific publications in organizational research. We focus on techniques of detecting hypothesis statements, classifying the causal and associative relationships in these statements, and deconstructing these relationships into entities and links. It is essential to distinguish associations and causal relationships, the latter of which is a stronger statement about the cause-and-effect logic (Pearl, 2009). It is crucial to detect and extract causal knowledge in organizational research, so that researchers and practitioners can draw evidence-based causation to design managerial and policy interventions.

Specifically, we developed machine reading models to complete three sequentially related tasks. The first task was ***hypothesis detection.*** We tried to identify whether a statement in a scholarly paper is a hypothesis or not, that is, whether this relationship was deliberately developed as a hypothesis for empirical testing. For this task, we used a model from the `fastText` *library*. `fastText` is an open-source library that does both word representations and text classification. This type of model has similar performance (e.g., accuracy, precision) as deep learning models but faster (Zolotov & Kung, 2017).

Our second task was ***relationship deconstruction***. Specifically, we deconstructed a hypothesis into cause entities, outcome entities. We used a two-layer stacked bi-directional *Long-Short Term Memory (*`LSTM`*)* architecture for the model, along with pre-trained GloVe word vectors (Pennington Socher, & Manning, 2014) for the text embeddings, which yielded good overall performance.

Our third task was ***feature classification*** (***causality and direction***). We classified a hypothesis as to whether it is stating a causal relationship or simply an association and classified the direction of the relationship in the hypothesis (positive, negative, nonlinear). We compared multiple models and pre-processing methods and found that the logistic regression model outperforms other methods. Furthermore, similar to prior works (e.g., Catalyst Team, 2016), we found that models using `bag-of-words` (`BOW`) features outperformed those using other features.

## MACHINE READING FOR LITERATURE REVIEWS

Machine reading for literature reviews is to engage NLP models to automate knowledge discovery and extraction from the scientific literature. As an emerging subfield of NLP, machine reading for literature reviews has been developed almost entirely in biomedical research, notably

Textpresso (Müller, Kenny, & Sternberg, 2004), GATE (Cunningham, Tablan, Roberts, & Bontcheva, 2013), Spá (Kuiper, Marshall, Wallace, & Swertz, 2014), and Reach (Valenzuela-Escárcega et al., 2018). These programs are built on pre-trained language representations of biomedicine, such as a taxonomy of biomedical entities (e.g., proteins) and events (e.g., biochemical interactions) of interest. Most machine reading models work on relatively simple jobs of extracting key findings from paper abstracts [for reviews, see, e.g., Marshall and Wallace (2019) and Jonnalagadda et al. (2015)]. As an exception, Reach (Reading and Assembling Contextual and Holistic mechanisms from text), recently developed by Valenzuela-Escárcega et al. (2018), adapts pre-trained NLP models to read full texts of biomedical databases, extracting biomedical entities (e.g., proteins) and the mechanisms linking these entities (e.g., "influences"). However, as *Reach* is built on biomedical taxonomies and corpus, it has limited application for social science papers.

Our approach follows the general principles of Reach and combines domain-specific rules and machine learning techniques to read the full texts of social science papers. Specifically, our approach takes four steps: Data preparation, hypothesis detection, relationship deconstruction, and relationship classification. Figure 1 illustrates the steps, which are discussed in detail in the following sections.

***Figure 1 about here***

## DATA PREPARATION

Our approach started with data collection for a sample textual data from publications (Section A in Figure 1). After extracting the hypothesis sentences and manually classifying them as explained above, we then randomly selected a relatively identical sample size of non-hypothesis statements from the same publications. As mentioned above, for hypothesis statements, we labeled each of the extracted sentences with four features: the cause, the outcome, the direction of the relationship, and whether this relationship is causal or not (causality). This labeling practice generally mimics the process of information reduction by human researchers. By reducing a large volume of publications into an annotated corpus, researchers can analyze and organize the four features to briefly understand the main findings of the literature.

**Collecting a Corpus**

To ground the NLP models into the domain of organizational research (Section A1 in Figure 1), we started by collecting a sample of papers related to organizational research in social sciences. We restrict our search of papers based on the explicit inclusion of organizatonal performance as part of the research question. In line with the new paradigm of multi-stakeholder and multi-dimensional conceptualization of corporate purpose (Harrison, Phillips, & Freeman, 2020), we defined organizational performance as an organization's effectiveness in meeting the expectations of two or more stakeholder groups (investors, employees, customers, and communities).

Based on the ISI Web of Science database of publications, all empirical publications (excluding meta-analysis) were first downloaded and read, as long as at least one keyword was directly suggesting a stakeholder group. The keywords indicating stakeholders were: *stakeholder\**, *investor\**, *shareholder\**, *owner\**, and *financ\** for investors; *customer\**, *consumer\**, and *user\** for consumers; *employee\**, *worker\**, *workforce\**, *labor\**, *labour\**, and *human resource\** for employees; and *communit\**, *societ\**, *environment\**, *climate\**, *natural resource\**, *responsib\**, and *social performance\** for the community. A snowball approach was adopted, in which each newly found performance construct will be added as a new keyword for the next search until no new construct was found. With the pool of papers collected above, we further shortlisted papers that included theoretical developments related to performance measures concerning at least two stakeholder groups. This sample represents high-quality scientific journal articles and offers a viable corpus of testable knowledge (i.e., hypotheses) concerning organizational performance.

The primary studies included two stakeholder groups for measuring organizational performance: the correlations between a factor and at least two stakeholder values. In total, we have identified and downloaded 138 peer-reviewed articles published between 1990 and 2018. We further removed 13 papers of which the PDFs were of poor quality for optical character recognition (OCR). The remaining 125 papers represent cross-disciplinary literature in social sciences in 1990-2018 to explain different organizational performance dimensions. The complete reference of these papers is listed in Supplementary materials S1.

**Developing a Sample for Hypothesis Detection**

Then we prepared this corpus for NLP model development (Section A2 in Figure 1). First, we converted each PDF (e.g., "paper.pdf") to raw text ("paper.txt"). We removed any tables, figures, and commonly used stop-words from articles (using the built-in dictionary by Python `NLTK` package). We then continued with developing an algorithm to identify which statements are likely hypotheses. Specifically, the algorithm works like the following. It detects any statements in a format similar to the following:

*"Hypothesis 1: ..."*

*"H1: ..."*

Specifically, we trained the algorithm to search for sentences that included targeted expressions like "Hypothesis" (or "Proposition") or "H" (or "P") followed by a number. This gave us 2,230 sentences that potentially contained hypotheses. We ended up with many false-positive extractions (i.e., sentences that contained the targeted expressions related to hypotheses but were, in fact, not the original hypothesis statements but explanations or simply mentioning of them). For instance, researchers often refer to a hypothesis when discussing the evidence. We screened all the 2,230 sentences manually and kept actual hypothesis statements. We ended up with 643 hypothesis statements across our 125 papers.

Below is an example of extracted hypothesis sentences:

*"H1. Commitment configuration is positively associated with firm performance."*

Finally, we constructed a relatively balanced corpus of 1,300 sentences by randomly drawing from the same publications 657 non-hypothesis sentences that also included "Hypothesis" (or "Proposition") or "H" (or "P") followed by a number. Essentially, we aimed to train classification models to distinguish the original hypotheses from the in-text mentions of them (e.g., discussion of empirical findings for a hypothesis).

**Annotating Features of Hypothesis Statements**

The next task was to develop models to extract information from each hypothesis statement. The objective was to reduce each hypothesis to its four key features: node 1 (a construct), node 2 (another construct), the direction of the link (positive, negative, or nonlinear), and the nature of this link (causal or associative statements). Below are two sets of examples that were classified as causal statements and association statements, respectively.

Examples of causal statements:

*"H1: The environmental legislation exerts a positive influence on the manager's perception about the environment as a competitive opportunity."*

*"H1: Stakeholder management will have a positive effect on CEO compensation levels."*

Examples of association statements:

*"H1: Stakeholder relations are negatively associated with the persistence of inferior financial performance."*

> *"H1: The grafting of new management team members after venture start-up is positively related to venture performance."*

We manually classified each hypothesis sentence into nodes, the direction of the link, and the nature of the link. We use these features as inputs to perform classification tasks later. Six well-trained graduate students in data science from an elite university completed the feature coding work. Each statement was coded by two different students independently. The inter-coder agreement was 95%, with the remaining disagreements fully resolved after a direct conversation. A co-author who specializes in organizational research played quality control to make sure the final coding was 100% correct. As an example, the last hypothesis statement cited earlier was annotated into the following features: Node 1 ("the grafting of new management team members after venture start-up"), Node 2 ("venture performance"), the direction of the link (positive), and the nature of the link (association). In total, we have manually completed these annotations for the 643 hypotheses that we extracted.

**A Summary of the Annotated Corpus**

The 643 hypothesis statements reported a mean of seven hypotheses per article and a standard deviation of five hypotheses. Typically, a set of hypothesis statements is one or three sentences long. As a reference, generally, an English sentence has on average 15 to 20 words (Plain English Campaign, 2004). Thus, we censored extractions by dropping sentences with more than 60 words, assuming they are not hypotheses in any organizational research papers. As Figure 2 illustrates, after this censoring, each hypothesis statement's number of words was approximately following a normal distribution, with a mean of 18.5 words and a standard deviation of 9.8 words. The data that support the findings of this study are available from the corresponding author upon reasonable request.

***Figure 2 about here***

## TASK 1: HYPOTHESIS DETECTION

After constructing the corpus, we develop text classification models to detect whether a sentence is a *hypothesis sentence* or not (Section B in Figure 1). As mentioned earlier, our corpus contained our final sample contained 1,300 sentences (including 643 hypothesis statements and randomly extracted 657 non-hypothesis sentences from the same sample of publications). This corpus was then divided for 10-fold cross-validation. Specifically, the whole sample was randomly split into ten subsamples. In each testing, nine subsamples (90% of the entire sample) were used as the training set to train the text classification models for identifying hypothesis statements. The remaining subsample (10% of the whole sample) was used to measure the training model's out-of-sample performance. We repeated this process ten times, in each of which we used a different 10% subsample as the test. We then reported the average out-of-sample performance as the overall performance of the training model. By averaging performance in ten sets of testing in different sets of subsamples, we would avoid overfitting bias. We also replicated the division to 75% training set and 25% test set and received highly consistent results.

Text classification models do not need to understand the meanings or grammatical structures within texts. Instead, we let statistical models predict the classification (1 for hypothesis and 0 for non-hypothesis). We fed text classification models with features of a sentence to find statistical relationships between features (inputs) of a raw sentence and the classification of this sentence (output). For example, if the word "associated" were more related to hypothesis sentences than non-hypotheses, the model would be more likely to classify sentences with the word "associated" as a *hypothesis* without knowing its meaning. Our sample met the requirement for successful text classification models requirement, as it covered a wide range of possible hypothesis-related words.

We used the text classification models from the `fastText` library – a supervised machine learning model –to classify sentences as a hypothesis or not (Section B1 in Figure 1). Facebook's AI Research group created this algorithm to learn word embeddings and perform text classification. This model has been shown to have similar performance (e.g., accuracy, precision, and F1-scores) as more complex deep learning models but at a significantly faster speed (Zolotov & Kung, 2017). Thus, it meets the purpose of our project, that is, saving time for research reviews.

Specifically, the algorithm of `fastText` model works as the following:

a) It breaks a sentence apart into separate tokens. Each token is a commonly used clause term or a word;

b) It assigns every token in the training sample an *n*-dimensional numerical vector (word embedding);

c) It assigns every sentence an *n*-dimensional numerical vector that averages the values of every dimension of the word's vectors in the sentence (sentence embedding);

d) The sentence embeddings are finally used as features (inputs) into a supervised classification model to predict the classification (hypothesis or non-hypothesis).

After comparing the preliminary performance of different linear and nonlinear supervised models, we used a neural network with one hidden layer and iterated through word- and sentence embeddings. The embedding for a given sentence and its associated label vector were very close to each other in a vector space. Finally, sentence embeddings were used as features for the final prediction.

*** Table 1 about here ***

We trained the `fastText` model after tunning model parameters (parametrization). Table

1 presents the four best-performing parametrizations after trying several different combinations in the values of the parameters. We find the order of words played no effect on the results of identifying hypothesis sentences. Furthermore, models using bi-grams, compared to those using uni-grams, reported a lower accuracy under all specifications. Also, the negative-sampling loss provided a better accuracy under most specifications. The best specification was Parametrization 4 in Table 1, which used uni-grams, a learning rate of 0.3, a 120-dimension vector to represent words, and the negative-sampling loss function. As presented in Table 1, we achieved an F-1 score of 96.7% for this specification on the test data, where the F-1 score is a comprehensive measure of model accuracy combining Precision and Recall. *Precision* is the ratio between the true positives (correctly predicted hypotheses) and all the positives (correctly and incorrectly predicted hypotheses), and *Recall* is the measure of our model correctly identifying true positives (percentage of correctly predicted hypotheses among all actual hypotheses).

F-1 score was calculated as:

$$F-1\ Score = \frac{2(precision * recall)}{(precision + recall)}$$

**Assessing the Interpretability of the Model**

One limitation of machine learning models is that they are often difficult to interpret. As a result, it cannot be trusted that these models have picked up the data's meaningful features. For instance, if hypothesis sentences are on average shorter (or longer) than non-hypothesis sentences in our sample, then the model might have classified short (or long) sentences as hypotheses and others as non-hypotheses. In this case, the model would report a high accuracy but is not based on meaningful features that define a hypothesis and thus may not perform effectively in new samples.

Ribeiro, Singh, and Guestrin (2016) introduced an approach to interpreting complex machine learning models, named Local Interpretable Model-Agnostic Explanations (`LIME`). Following `LIME`, we need to explain how the `fastText` model predicts by training a simpler stand-in model, then use this simpler stand-in model to explain the original `fastText` model's prediction (Section B2 in Figure 1). Even though the simpler model cannot capture all of the `fastText` model's complexity, it helps to understand the logic the complex model might have used. Instead of training the stand-in model on the entire sample, we used a subsample of the data for the stand-in model to classify one sentence correctly. As long as the stand-in model used the same logic as the *fastText* model, we would understand and explain the predictions made by `fastText`.

To construct the stand-in model's training set, we created many variations out of each sentence, each time removing specific words. In hypothesis detection, we classified a hypothesis sentence multiple times by removing a different word each time from the sentence. In this way, we estimated each word's relative importance in the final prediction. By making several predictions for many variations of the same sentence using `fastText` (i.e., missing different words), we were essentially capturing how the model weighted different words as a way of "understanding" that sentence. Finally, we used the sentence variations and classification predictions as the training set to train the stand-in model using the Simple Linear Classification Model.

We want to note that this approach's shortcoming is the implicit focus on only the importance of single words, not phrases or *n*-grams. However, as we will show, this limitation does not prevent us from making reasonable interpretations of `fastText`. Specifically, our stand-in model's outputs were the weights assigned to each word in the hypothesis sentence, where the weights represent how much that word affected the final prediction.

*** Figures 3 and 4 about here ***

Figure 3 shows that the words "positively" and "associated" were among the most important words as they contributed the most to the classification of a sentence as a hypothesis. Figure 4 shows that the words contributing the most to classifying a sentence as a *non-hypothesis* were "significant" and "regression." They are usually not part of the original hypothesis sentence but were used to discuss the empirical test for or against the hypothesis. However, no word in this sentence was strongly associated with a hypothesis sentence. Therefore, from these two figures, it seems clear that the `fastText` model was valuing the correct words to make predictions regarding hypothesis detection.

**TASK 2: RELATIONSHIP DECONSTRUCTION**

We then developed our NLP model to extract the key features in a relationship from each hypothesis, including two nodes (constructs) and the link between them from a sentence (Section C in Figure 1). For example, if we have an association statement, "*Node1 is related to Node2,*" we want to extract both "Node1" and "Node2." But if we have a causal statement, "*Node 1 causes Node 2,*" then we need to not only extract "Node1" and "Node 2" but also identify "Node 1" as the cause and "Node 2" as the outcome.

First, we labeled the nodes in our sample data. For each hypothesis sentence, we labeled non-nodes as "0", the "cause" node as "1", and the "outcome" node as "2". In the case of atypical hypotheses such as more than two nodes (e.g., multiple causes or outcomes) and more than one link (e.g., moderators), we aggregated multiple nodes of the same level together to form a Node 1- link- Node 2 structure. Specifically, in the case of more than two nodes, such as "A would reduce B and C," we treated "A" as Node 1 and "B and C" together as Node 2. In the case of multiple links, such as "A is moderating the relationship between B and C," we treated "A" as Node 1 and "the relationship between B and C" together as Node 2.

Again, six well-trained graduate students in data science completed the feature coding work. Each statement was coded by two different students independently. The inter-coder agreement was 90%, with the remaining disagreements fully resolved after a direct conversation. A co-author who specializes in organizational research played quality control to ensure the final coding was 100% correct.

We padded each of the sentences, so they were formatted to have the same dimension of 50 (i.e., the vector dimension). We then fitted the data to a model with the following architecture listed:

a) Text vectorization layer, which standardizes each text and utilizes only uni-grams;
b) Embedding layer, which applies the pre-trained words vectors based on `GloVe`;
c) One dimensional spatial dropout layer with a dropout rate of 0.5;
d) Two-layer stacked bi-directional `LSTM`, with 32 units on the first layer, and 128 units on the second, both with a recurrent dropout rate of 0.1;
e) Time-distributed dense output layer;

Besides, we use the `RMSprop` back-propagation optimizer, with loss calculated by categorical cross-entropy. Complete visualization of the model can be seen in Appendix 1, generated via `Net2Viz` (Alex Bäuerle & Timo Ropinski, 2019).

** Figure 5 and Table 2 about here**

We ran the model with a batch size of 32 and 50 epochs to minimize training overfitting. Figure 5 shows the training and test accuracy over the number of epochs. We received a very high accuracy of 97.2% from the testing data, measuring the total percentage of true positives (correctly predicted nodes and links) and true negatives (correctly predicted non-nodes and non-links). However, the dataset is highly imbalanced, with approximately 90% of all tokens representing non-node or non-link entities, 5% representing cause entities, and 5% representing outcome entities. Thus, accuracy may be an inappropriate indicator of model performance, and we need to rely on additional performance metrics, including precision, recall, and F1-score, to evaluate the model. Table 6 shows the additional metrics on different predictions, all of which are satisfactory – significantly over 90% in all measures.

## TASK 3: FEATURE CLASSIFICATION

**Classifying the Nature of the Link (Causality or Association)**

We moved on to develop a model to classify if a sentence made a causal statement or not (Section D1 in Figure 1). We created two different representations from each hypothesis. The first representation was word embedding based on `BOW` features. Specifically, we identified the frequency of uni-gram, bi-gram, and tri-grams against the complete corpus (1,300 sentences). The second representation was a sentence embedding using `Doc2Vec` (`D2V`). With these two different representations, multiple classification models were evaluated. For both `BOW` and `D2V` features, we used and evaluated the following classification models: logistic regression, random forest, and support vector machine (SVM). We also used synthetic oversampling methods `SMOTE` and `ADASYN`, which did not exhibit any significant model improvements. Thus, there was no merit for the use of synthetic data.

***Table 3 and 4 about here***

Prediction performance metrics of different classification models with `BOW` and `D2V` are reported in Tables 3 and 4, respectively. Models using `BOW` representation generally performed better than `D2V` representation. Among all evaluations, logistic regression using `BOW` features produced the greatest F1-score. We further tuned the hyperparameters on this model, using stratified 10-fold cross-validations and three repeats. This hyperparameter tuning yielded a further improved F1-score as high as 92.4% (see Table 3).

**Classifying the Direction of the Link (Positive, Negative, or Nonlinear)**

***Tables 5 and 6 about here***

We then trained a model to classify the direction of the link in a hypothesis (positive, negative, or nonlinear) (Section D2 in Figure 1). This process used the same feature representations (`BOW` and `D2V`), models, and oversampling methods as the feature classification model. Prediction performance metrics for different classification models with `BOW` and `D2V` are reported in Tables 5 and 6, respectively.

Models using `BOW` representation generally performed better than `D2V` representations. Logistic regression using `BOW` features produced the greatest F1-score. We further tuned the hyperparameters on this model, using stratified 10-fold cross-validations and three repeats. This hyperparameter tuning yielded a further improved F1-score as high as 85.9% (see Table 3).

**A USER'S GUIDE**

In this project, we constructed an interdisciplinary corpus of hypothesis statements from a set of high-quality peer-reviewed papers in social sciences. Then we used this data to train models that perform three different tasks that mimic how human researchers typically extract theoretical insights from the literature for research reviews. We recognize that most organizational researchers would be direct users of the existing pre-trained models for machine-reading, rather than those who have the programming background to re-train the models for a different task. For this large audience, we have made several efforts to make our models fully accessible, that is, a simple drag-and-drop with minimum coding. First, we have developed a free R Shiny app as the graphic user interface (GUI). On this GUI, users can upload an unlimited volume of papers as PDFs to initiate the pre-trained models to automatically parse texts into a corpus and then play all three tasks. Second, we have connected the R Shiny app through `r-reticulate` package to convert the Python programs into R programs. The R Shiny app then runs on both R and Python programs on the back end.

Now we illustrate how users without a programming background can install and use our pre-trained models via an R Shiny app in detail. We have developed the R package `HypothesisReader` and stored it on Github for remote installations. The package implements the methodology outlined in this paper and automatically launches the pre-trained models for users' own PDF data. The following software should be pre-installed in a user's computer.

a) Java 8 or OpenJDK 1.8
b) R and R package `"devtools"`

**Installation Steps**

a) Open R and install R package from GitHub repository by typing the following:

   `devtools::install_github("canfielder/HypothesisReader")`

   When prompted `"Enter one or more numbers, or an empty line to skip updates:"`, simply hit the Enter key;

b) Execute the function below to launch the R Shiny app GUI:

   `HypothesisReader::LaunchApp()`

c) Upload PDFs on the GUI to initiate the text processing and install Python package;

d) At the prompt in the console, select *y* to install Miniconda;

e) Restart R session (Session > Restart);

f) The pre-trained models are now ready for use.

**Troubleshooting**

If any of the required Python packages do not automatically install (which would yield an error), installation can be forced with the following function in R:

```
HypothesisReader:: InstallHypothesisReader()
```

**Usage**

Finally, we provide a stepwise illustration of using our pre-trained models via an R Shiny app GUI. As shown in Appendix 2, using the tool takes three simple steps: a) launch the GUI through R, b) upload the PDF data, and c) download the deconstructed data in CSV.

## DISCUSSION

We suggest several directions of future research are valuable for improving our models. First, our models currently force each hypothesis into a three-part structure – two nodes and one link. The majority (82%) of the hypotheses in our sample follow this structure to contain two separate constructs. However, there are exceptions, such as moderators and multiple causes or outcomes. Currently, our models would aggregate nodes or links at the same level to force a hypothesis into three parts. Such cases include a) more than two nodes or b) more than two links (moderators and, in rare cases, mediators). As an example for more than two nodes, our sample contains the following hypothesis with multiple outcomes: "*increased use of high-performance work systems results in increased labor productivity, increased workforce innovation, and decreased voluntary employee turnover.*" Currently, our pre-trained models would deconstruct it into Node 1 ("*increased use of high-performance work systems*"), Node 2 ("*increased labor productivity, increased workforce innovation, and decreased voluntary employee turnover*"), and a link (nature=positive; causality=1). However, the ideal outputs should be three relationships with a shared Node 1 as "*use of high-performance work systems*.": a) Node 2 as "*labor productivity*" with a link (nature=positive; causality=1); b) Node 2 as "*workforce innovation*" with a link (nature=positive; causality=1); c) Node 2 as "*voluntary employee turnover*" with a link (nature=negative; causality=1).

As an example for more than two links, our sample contains some hypotheses on moderating effects like "*the positive relationship between corporate philanthropy and a firm's financial performance increases with its advertising intensity*." Currently, our pre-trained models would deconstruct this relationship into Node 1 ("*the positive relationship between corporate philanthropy and a firm's financial performance*"), Node 2 ("*advertising intensity*"), and a link (nature=positive; causality=0). The ideal outputs should generate an additional relationship with Node 1 as "*corporate philanthropy*," Node 2 as "*a firm's financial performance*," and a link (nature=positive; causality=0). As another example for more than two links, our sample contains hypotheses that combine two causal relationships through a mediating process, such as "*marketing competence mediates the relationship between CSR toward society and firm performance*." Currently, our pre-trained models would deconstruct this relationship into Node 1 ("*marketing competence*"), Node 2 ("*the relationship between CSR toward society and firm performance*"), and a link (nature=nonlinear; causality=1). However, the ideal outputs should divide this relationship into two causal relationships. The first relationship should have Node 1 as "*CSR toward society,*" Node 2 as "*marketing competence,*" and a link (nature=positive; causality=1). The second relationship should have Node 1 as "*marketing competence,*" Node 2 as "*firm performance,*" and a link (nature=positive; causality=1).

Currently, our training is limited by the small sample of such exceptional cases. We propose to increase the size of our sample by annotating a more extensive corpus that contains significantly more atypical hypotheses, including more than two nodes, moderators, and mediators. A larger sample would also significantly improve the training and the out-of-the-sample performance.

Second, we suggest future studies should also develop clustering models to sort and aggregate extracted nodes into a standardized taxonomic hierarchy. For instance, after deconstruction, our sample contains expressions of nodes like "CSR towards society," "social performance," and "social responsibility." Currently, the outputs would export the original forms of each, and thus would treat them as different constructs. We propose to develop a standardized taxonomy of commonly used terms in organizational research to sort and aggregate semantically similar constructs into the same new construct. For instance, the three mentioned examples could be grouped into a new construct called "firm performance towards the society." As the literature continues to evolve and grow, a challenge is that many constructs may be introduced to the field without precisely fitting into an existing taxonomy. We suggest a highly valuable approach would be to use unsupervised learning to cluster contructs automatically without a pre-defined taxonomy. We suggest that researchers draw a larger corpus of research documents such as company reports, Wikipedia, and textbooks to triangulate each construct's semantically adjacent words (e.g., N-grams) and use adjacent words to cluster constructs together.

Finally, we suggest that researchers with advanced NLP training can further refine our methodology and re-train our models for different tasks. Currently, our approach applies only to hypotheses, that is, testable theoretical statements. As literature reviews are often accompanied by empirical syntheses such as meta-analysis and meta-regressions, researchers often would like to detect and extract the empirical findings. Researchers could go beyond hypotheses and focus on detecting and comparing empirical evidence by focusing on a different set of trigger words. Rather than using only "Hypothesis" (or "Propositions) or "H" (or "P") followed by a number, we could combine them and with other trigger words indicating empirical evidence, such as "support,"

"supportive," "evidence," "significant," and so on. In this way, we could train models to detect empirical findings and classify each hypothesis as "supported" or "unsupported." This, however, would be more challenging to develop, as not all empirical evidence is mentioned in the text. Many empirical details, such as coefficients and p-values, are only reported in Tables without specific mentions in the paper. However, for meta-analytic reviews, it would also require that the machine reading models extract the same information in papers where a focal variable was tested only as a control variable and thus unmentioned specifically as hypotheses anywhere in the paper.

**Table 1. Evaluation of Hypothesis Detection Models**

| Model | N-grams | Learning Rate | F1-score | |
|---|---|---|---|---|
| | | | SoftMax | Neg Sampling |
| Parametrization 1 | 1 | 0.1 | 87.10% | 92.60% |
| Parametrization 2 | 2 | 0.1 | 84.60% | 85.70% |
| Parametrization 3 | 5 | 0.1 | 85.10% | 55.30% |
| Parametrization 4 | 1 | 0.3 | 95.70% | 96.70% |

Note: We used 120-dimensional vectors. F1-Score was calculated using two loss functions: Soft Max and Negative Sampling. We used word N-grams (N=1, 2, and 5).

**Table 2. Evaluation of Relationship Deconstruction Models**

|  | Precision | Recall | F1-Score |
|---|---|---|---|
| Overall (All Nodes) | 92.4% | 91.9% | 92.2% |
| Non-Label (0) | 98.6% | 98.7% | 98.6% |
| Cause (1) | 88.8% | 89.9% | 89.4% |
| Outcome (2) | 89.8% | 87.2% | 88.5% |

**Table 3. Evaluation of Models using BOW Features to Classify the Nature of the Link**

| Model | Feature Normalization | Accuracy | Precision | Recall | F1-Score |
|---|---|---|---|---|---|
| Logistic Regression* | Stemming | 93.7% | 93.5% | 91.4% | 92.4% |
| Random Forest | Lemmatization | 90.6% | 94.0% | 84.4% | 87.6% |
| Support Vector Machines | Stemming | 93.1% | 92.5% | 90.9% | 91.6% |

* Model with the greatest F1-score as the overall performance measure.

**Table 4. Evaluation of Models using D2V Features to Classify the Nature of the Link**

| Model | Feature Normalization | Accuracy | Precision | Recall | F1-Score |
|---|---|---|---|---|---|
| Logistic Regression | Stemming | 73.6% | 68.7% | 62.2% | 63.0% |
| Random Forest* | Lemmatization | 77.4% | 78.5% | 64.9% | 66.3% |
| Support Vector Machines | Lemmatization | 70.4% | 85.1% | 51.0% | 43.3% |

* Model with the greatest F1-score as the overall performance measure.

**Table 5. Evaluation of Models using BOW Features to Classify the Direction of the Link**

| Model | Feature Normalization | Accuracy | Precision | Recall | F1-Score |
|---|---|---|---|---|---|
| Logistic Regression* | Stemming | 91.3% | 87.6% | 84.6% | 85.9% |
| Random Forest | Stemming | 85.7% | 89.3% | 67.5% | 72.0% |
| Support Vector Machines | Stemming | 85.7% | 80.9% | 70.7% | 74.4% |

* Model with the greatest F1-score as the overall performance measure.

**Table 6. Evaluation of Models using D2V Features to Classify the Direction of the Link**

| Model | Feature Normalization | Accuracy | Precision | Recall | F1-Score |
|---|---|---|---|---|---|
| Logistic Regression* | Lemmatization | 70.2% | 47.4% | 38.1% | 38.7% |
| Random Forest | Stemming | 73.9% | 58.2% | 35.7% | 33.6% |
| Support Vector Machines | Lemmatization | 73.9% | 24.6% | 33.3% | 28.3% |

* Model with the greatest F1-score as the overall performance measure.

**Figure 1. Overview of Methodological Approach**

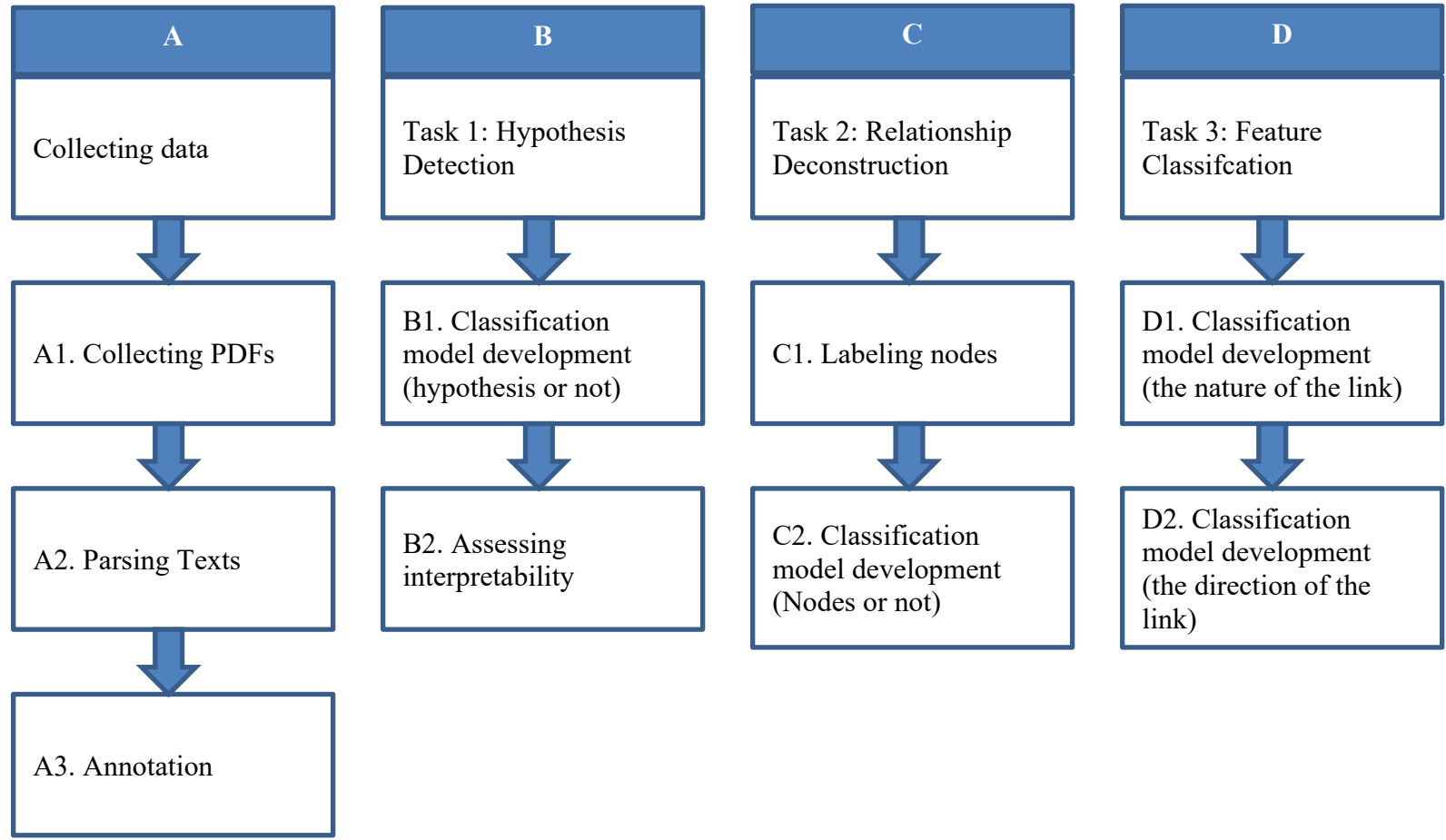

**Figure 2. Number of Words per Sentence in Our Corpus**

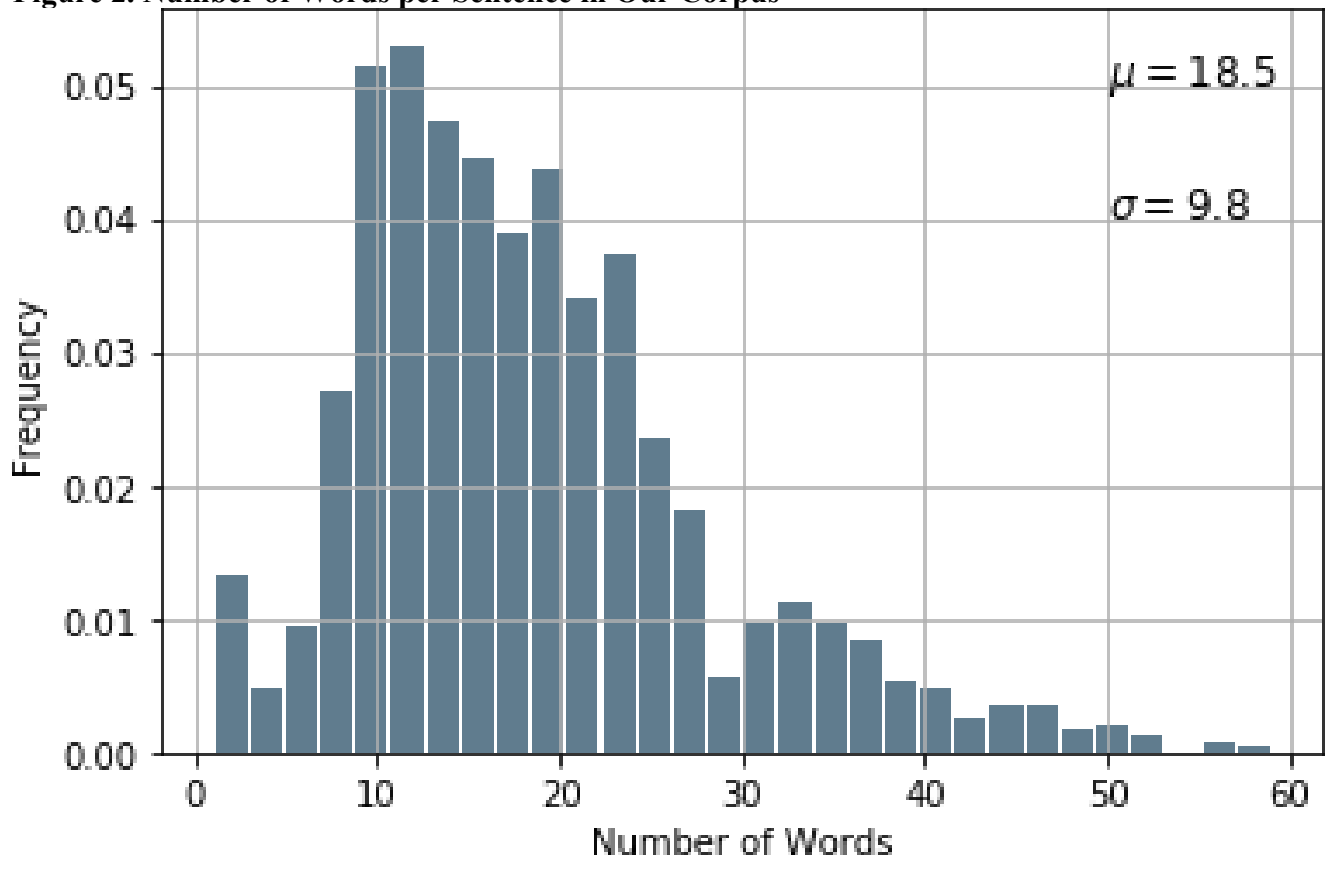

**Figure 3. An Example of Hypothesis Sentence**

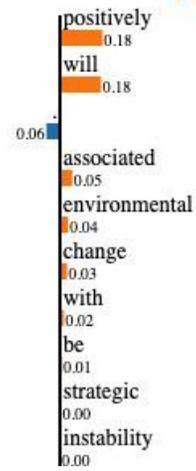

**Figure 4. An Example for a Non-Hypothesis Sentence**

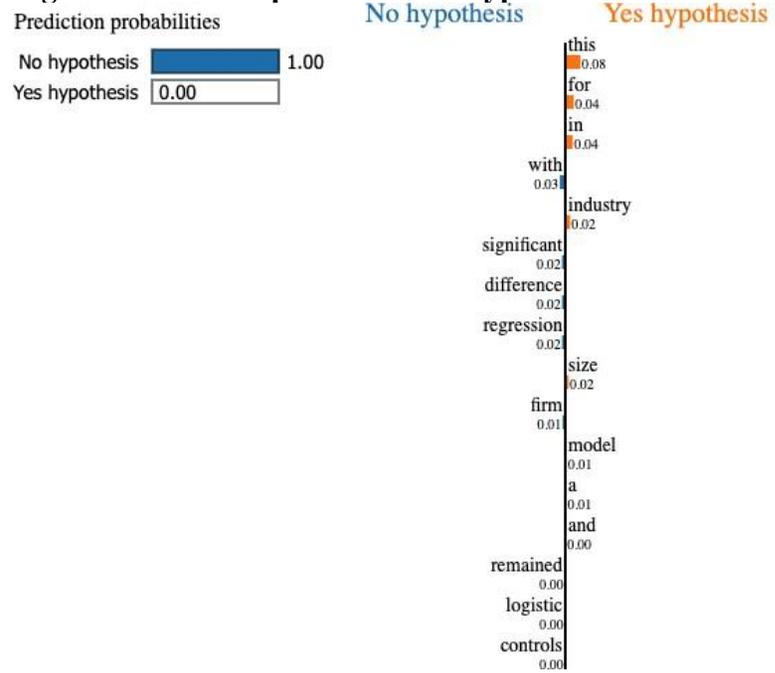
<-</->

**Figure 5. Performance of Relationship Deconstruction**

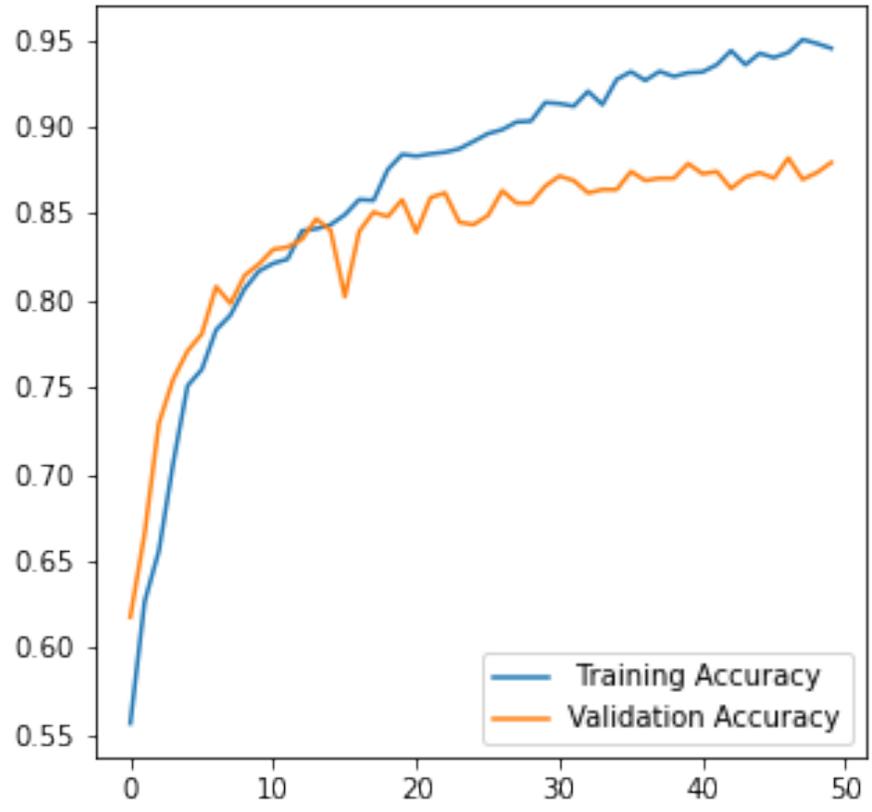

**Appendix 1. Relationship Deconstruction Model Structure**

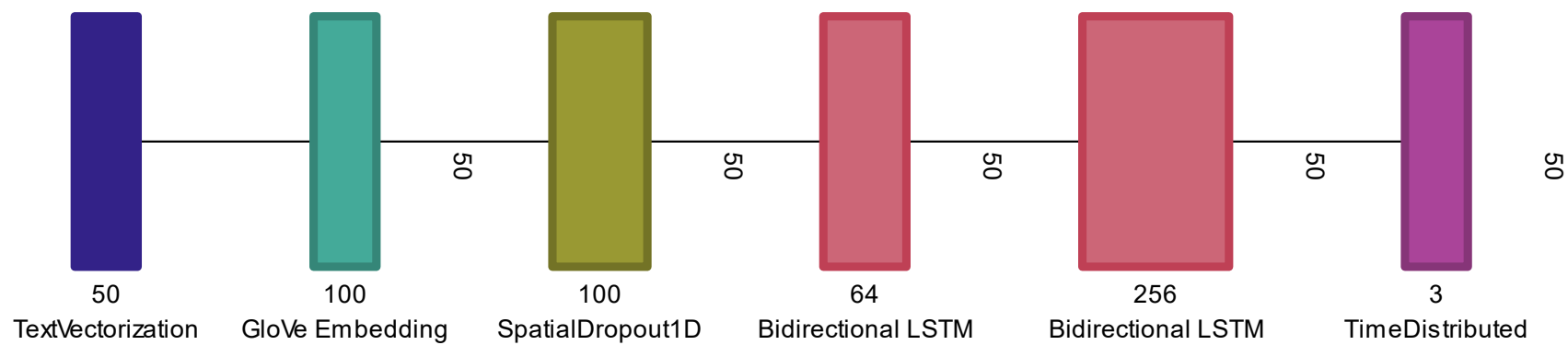

**Appendix 2. Usage of Pre-trained Models via R Shiny App**

Step 1. Launch Pre-trained Models via R Shiny GUI

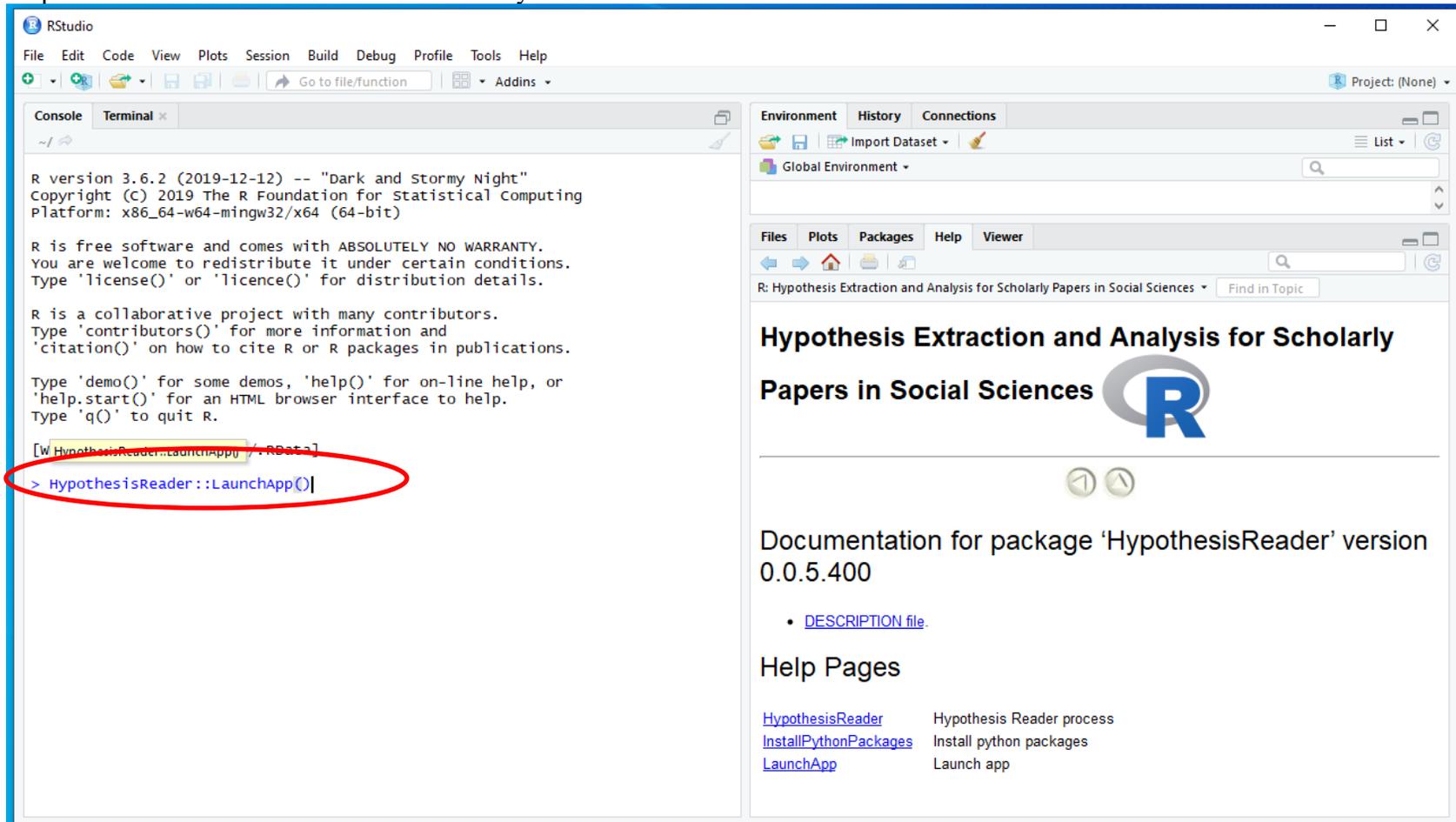

Step 2. Upload all PDFs by clicking the Browse button

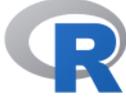

Step 3. Download the Deconstructed Data of Hypotheses

| file_name | hypothesis_num | hypothesis | variable_1 | variable_2 | direction | causal_relationship |
|---|---|---|---|---|---|---|
| 1 bc11amj.pdf | h_1 | work organization practices that enhance employee discretion and group collaboration will be associated with lower quit rates and lower dismissal rates | work organization practices that enhance employee discretion and group collaboration | quit rates and lower dismissal rates | neg | 0 |
| 2 bc11amj.pdf | h_2 | employment practices emphasizing inducements and investments will be associated with lower quit rates and lower dismissal rates | employment practices emphasizing inducements and investments | quit rates and lower dismissal rates | neg | 0 |
| 3 bc11amj.pdf | h_3 | performance-enhancing practices will be positively related to both quit rates and dismissal rates | performance-enhancing practices | quit rates and dismissal rates | pos | 0 |
| 4 bc11amj.pdf | h_4a | high involvement work organization, investment and inducement practices, and performance-enhancing expectations will each individually be associated with higher levels of operational performance | high involvement work organization, investment and inducement practices, and performance-enhancing expectations | operational performance | pos | 0 |
| 5 bc11amj.pdf | h_4b | the interactions of high involvement work, investment and inducement practices, and performance-enhancing expectations will be associated with higher levels of operational performance than the simple additive effect of these hr practices | interactions of high involvement work, investment and inducement practices, and performance-enhancing expectations | operational performance | pos | 0 |
| 6 bc11amj.pdf | h_5a | quit and dismissal rates will be negatively related to operational performance: these theoretical arguments also suggest that whether turnover mediates the relationship between hr practices and performance may depend upon the contingencies discussed above | quit and dismissal rates | operational performance | neg | 0 |
| 7 bc11amj.pdf | h_5b | the hr performance relationship will be mediated by the additive effect of quit rates and dismissal rates | effect of quit rates and dismissal rates | hr performance | non_lin | 1 |

Showing 1 to 7 of 7 entries

⬇ Download

# SUPPLEMENTARY MATERIALS

**S1 Studies included in the corpus**